\documentclass{aa}
\usepackage{epsfig}
\usepackage{amssymb}

\begin{document}
\thesaurus{03(08.06.2, 08.16.5, 08.05.10} 
\title{On the Formation of Massive Stars by Accretion }

\author{Peder Norberg \inst{1} \and Andr\'e Maeder\inst{2}}
   
   \offprints{A. Maeder}

   \institute{ Department of Physics, University of Durham, 
Durham DH1 3LE, UK \and Observatoire de Gen\`eve, 
              CH-1290 Sauverny,
	      Switzerland}

   \maketitle
   
   \markboth{P. Norberg \& A. Maeder: Massive Star
   Formation}{P. Norberg \& A. Maeder: Massive Star Formation}

\begin{abstract}

At present, there are two scenarios for the formation of massive stars: 
1) The accretion scenario and 2) The coalescence scenario, which implies
the merging of intermediate mass stars. We examine here some
properties of the first one. Radio and IR observations by Churchwell (\cite{Chu99}) and Henning et al. (\cite{Henn00})
 of mass outflows around massive 
Pre-Main Sequence (PMS) stars 
show an increase by several orders of magnitudes 
of the outflow rates with stellar luminosities, and thus
with stellar masses. As typically, a fraction of $\frac{1}{3}$ to
$\frac{1}{6}$ of the infalling material is estimated to be accreted,
this suggests that the accretion rate is also quickly increasing with the stellar mass.

We calculate three different sets of birthlines, i.e. tracks followed
by a continuously accreting star. First, three models with a constant
accretion rate ( $\dot{M}_{\rm{accr}}$ = $10^{-6}$, $10^{-5}$,
$10^{-4}$  M$_{\odot}$ yr$^{-1}$). Then several birthlines following
the accretion models of Bernasconi and Maeder (\cite{BM96}), which
have $\dot{M}_{\rm{accr}}$ increasing 
only slightly with mass. Finally we calculate several birthlines for which $\dot{M}_{accr} = \dot{M}_{\mathrm{ref}} \left( {\frac{M}{M_{\odot}}} 
\right) ^{\varphi}$, with values of $\varphi$ equal 
to 0.5, 1.0 and 1.5 and also for different values of $\dot{M}_{\mathrm{ref}}$. The best fit to the observations of PMS stars in the HR diagram is achieved for $\varphi$ between 1.0 or 1.5 and for $\dot{M}_{\mathrm{ref}} \simeq   10^{-5}$  M$_{\odot}$ yr$^{-1}$. Considerations on the lifetimes favour values of $\varphi$ equal to 1.5. These accretion rates do well correspond to those derived from radio and IR observations of mass outflows. Moreover they also lie in the ``permitted region'' of the dynamical models given by Wolfire and Cassinelli (\cite {WC87}).

We emphasize the importance of the accretion scenario for shaping the IMF, and in particular for determining the upper mass limit of stars. In the accretion scenario, this upper mass limit will be given by the mass for which the accretion rate is such that the accretion induced shock luminosity is of the order of the Eddington luminosity.


\end{abstract}

\section{Introduction}

The formation of massive stars is still a very uncertain domain of
stellar astrophysics. Schematically, there are at present two very
different scenarios. 1) \textit{The coalescence scenario} proposed by
Bonnell et al. (\cite{Bo98}) and Stahler et al. (\cite{Sta98}). The
formation of massive stars ($M \geq 10 M_{\odot}$) is assumed to occur 
by coalescence of stars of intermediate masses, which form through accretion
onto initially lower mass protostars. The basic reason for 
the development of this formation scenario was
the difficulty of accreting mass onto very luminous stars. 2) \textit{
The accretion scenario}  was initially proposed by Stahler et al. (\cite{Sta80a}, \cite{Sta80b}, \cite{Sta81}) and was further
developed for low and intermediate mass stars (see for example
Palla and Stahler,
\cite{PS93}). It was  then investigated as a formation mechanism 
for massive stars  by Beech and Mitalas (\cite{BeeM94}),
Bernasconi and Maeder (\cite{BM96}).  In this scenario,
the massive stars no longer cross horizontally the HR diagram, coming from
the red to the blue, on the Kelvin-Helmholtz timescale, but rise upwards
in the HR diagram along the so-called birthline. The birthline  is 
defined  as the path in the HR diagram followed by a continuously
accreting star. For low and intermediate mass stars,
the birthline forms an upper envelope of individual
evolutionary tracks in the HR diagram. The location of the birthline
and the timescales on it strongly depend on the accretion rates
$\dot{M}_{\rm{accr}}$ (Bernasconi and Maeder, \cite{BM96}; Tout et al.
\cite{Tout99}). Thus, in this
scenario it is very important to know how the accretion rate varies
with the mass already accreted onto the star.

Both scenarios have their own advantages and difficulties. They both influence 
the upper limit of stellar masses, but the physical mechanism determining
this limit is of course different for each of the two above scenarios.
In this paper we shall 
examine some properties of the accretion scenario, in order to
provide further arguments in the debate. By $\dot{M}_{\rm{accr}}$, it is usually meant the accretion rate onto the central body, and this is our adopted viewpoint throughout this whole paper. In further more complete models, we will distinguish between the accretion from the  molecular cloud to the disk and the accretion from the disk to the central protostar.

In Section 2, we examine some recent results on mass accretion and
outflows in ultra-compact HII (UC HII) regions. In Section 3, we
compare to the observations of pre-main sequence (PMS) stars
some  standard birthlines, i.e. with constant or slowly variable
$\dot{M}_{\rm{accr}}$. In Section 4, we calculate new birthlines
with quickly increasing accretion rates. In Section 5, we briefly give our conclusions, and
in particular we discuss the issue of the maximum stellar mass in the accretion paradigm.

\section{The accretion scenario for massive stars}

The accretion scenario, despite its successes 
(Palla and Stahler, \cite{PS93}), has been considered
to be impossible for massive stars, due to their high luminosities,
which are able to reverse the collapse (Bonnell et al. \cite{Bo98};
Stahler et al. \cite{Sta98}). In this case, the radiation pressure on the dust
is high enough, so that its momentum can be transferred to the gas
(Wolfire and Cassinelli, \cite{WC87}).
In the accretion scenario, the accretion rate $\dot{M}_{\rm{accr}}$
is an essential parameter, since it determines the momentum of the infalling
material.

 In their  discussion on the study of the difficulties to
form massive stars with the accretion scenario, Stahler et al. 
(\cite{Sta98}) assume that the accretion rate behaves like

\begin{eqnarray}
\dot{M}_{accr} \simeq  \frac{{c_{s}}^3}{G}
\label{acc_rate}
\end{eqnarray}

\noindent
where $c_{s}$ is the sound speed in the molecular cloud.
 They stress that a remarkable
property of this rate is that it is independent of the density of the
parent cloud. The value of $c_{s}$, of course, depends very much 
on the temperature of the cloud. 
Stahler et al. (\cite{Sta98}) assume that the pre-collapse
temperature is independent of the core density and mass, and also
suggest that this does not change too much, if the infalling material
goes via a disk instead of landing directly on the stellar 
surface.  With the assumption that clouds where massive stars form have
the same typical temperature  T = 10-20 K as those where low mass stars form, Stahler et al. (\cite{Sta98}) get accretion rates of $10^{-5}$ to $10^{-6}$ M$_{\odot}$ yr$^{-1}$. 
It is with this kind of assumptions that the constant accretion rates models 
by Beech and Mitalas (\cite{BeeM94}), and those with slowly varying accretion rates by Bernasconi and Maeder(\cite{BM96}) have been constructed.
The birthline of these models joins the zero-age sequence when 
the heat released by the nuclear reactions stops the stellar contraction.
This occurs typically around 8 to 10 M$_{\odot}$.

\begin{figure}[tb]
  \resizebox{\hsize}{!}{\includegraphics{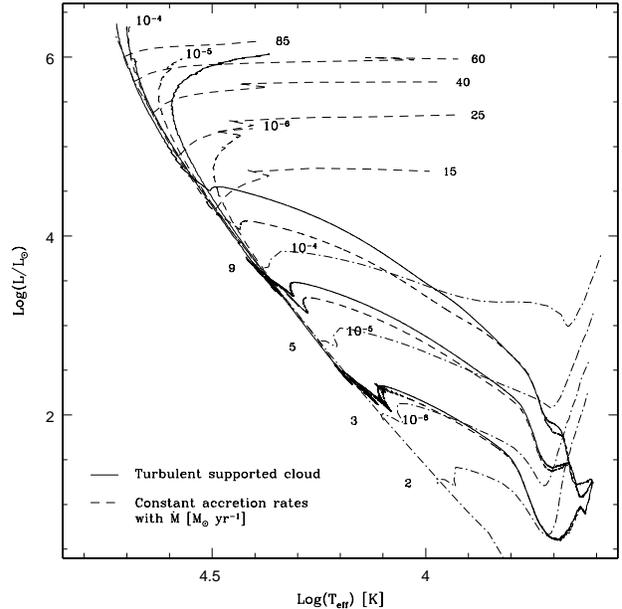}}
  \caption{Comparison between birthlines calculated with
constant accretion rates and those calculated with expression
3.
applied to the models by Bernasconi and Maeder (\cite{BM96}). The models are
given for $F$ = 0.1, 1.0 and 10.0 and they correspond to about
 $\dot{\mathrm{M}}_{cst} =$ 10$^{-6}$, 
 10$^{-5}$ and 10$^{-4}$ M$_{\odot}$ yr$^{-1}$. The dot-broken lines are the PMS tracks for constant mass with the
indicated value (Bernasconi and Maeder, \cite{BM96}).
The tracks with broken lines in the upper part are post-Main Sequence (post-MS) tracks
for massive stars of 15 to 85  M$_{\odot}$ by Schaller 
et al. (\cite{schaller}).}
  \label{comp}
\end{figure}

\begin{figure*}[tb]
  \resizebox{14cm}{!}{\includegraphics{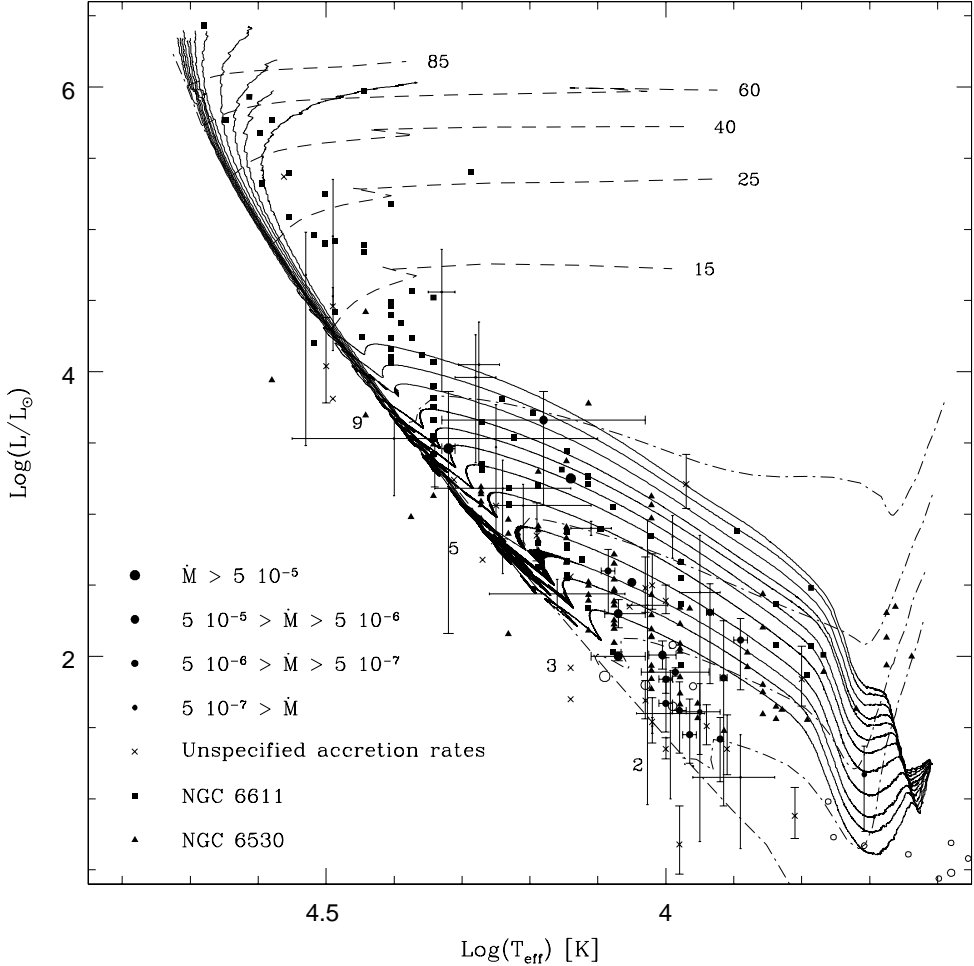}}
  \caption{This grid of birthlines (continuous lines) is made out of twelve tracks with the following values of $F$ =
 0.1, 0.15, 0.2, 0.3, 0.5, 0.75, 1.0, 1.25, 1.75, 2.5, 3.5, 5.0. In general, the luminosity increases with $F$ for a fixed value of the effective temperature. Pre-main sequence evolution tracks 
with constant mass are in dot-broken lines (Bernasconi and Maeder, \cite{BM96})
and the post-main sequence tracks from Schaller et al. (\cite{schaller}) are in short dashed lines. The labels correspond to the mass along each track. Observations are issued from: 1. compilation done by Bernasconi and Maeder (\cite{BM96}); 2. Hillenbrand et al. (\cite{hill92}); 3. Damiani et al. (\cite{dam94}); 4. Cohen and Kuhi (\cite{coh79}); 5. van den Ancker et al. (\cite{anck97a}); 6. Berrilli et al. (\cite{ber92}); 7. de Winter et al. (\cite{win97}); 8. Th\'e et al. (\cite{the90}); 9. van den Ancker et al. (\cite{anck98}); 10. van den Ancker et al. (\cite{anck97b}). For those stars 
with an error estimate on the luminosity and/or the effective temperature, 
the  error bars are plotted.
Moreover if several measures exist for a same star, we give the average value and indicate the existing dispersion on it. 
The value of  $\dot{M}_{\rm{accr}}$ is given by the size of the symbol, and an open symbol signifies that several groups have measured 
L and T$_{\rm{eff}}$ for this star.}
  \label{birth}
\end{figure*}

For massive stars with accretion rates of the order of  $10^{-5}$
M$_{\odot}$ yr$^{-1}$, it is true that the  momentum of 
the infalling material is
much smaller than the outwards radiation momentum of the star, 
which is thus able to reverse the accretion process. 
Therefore such low mass accretion rates are impossible for massive
stars. This is by the way also confirmed by the existence, around
these massive stars, of stellar winds of similar magnitudes but blowing 
in the opposite direction.
Moreover, we note that for this kind of $\dot{M}_{\rm{accr}}$ the 
formation lifetimes would be longer than 
the main sequence lifetime! 
 The dynamics of the infalling material on protostars
has been studied in detail by Wolfire and Cassinelli (\cite{WC87}).
They found that the abundance of dust as well as the size of the grains
should be reduced to allow infall. They examined the permitted
regions in a plane $\log \dot{M}_{\rm{accr}}$ vs. $\log M$, where $M$
is the mass
of the newly formed star. E.g. for a star of 40 M$_{\odot}$,
the allowed region lies between $\dot{M}_{\rm{accr}} \simeq 10^{-4}$
and $10^{-2}$ M$_{\odot}$ yr$^{-1}$.
 For lower $\dot{M}_{\rm{accr}}$, the accretion
is halted by the radiation field, while for higher $\dot{M}_{\rm{accr}}$
the total luminosity (the protostellar luminosity and the accretion 
induced shock luminosity) exceeds the Eddington luminosity.

There should be some relation between the accretion rates from
collapsing clouds and their temperature  $T$ 
(Wolfire and Cassinelli (\cite{WC87})). However, 
this is true only if the thermal support is the only 
source of  support in the clouds. In this case, which is likely 
not realistic, the temperature  of the  collapsing clouds
  leading to
massive stars should be as  high as T $\geq$ 200 K and maybe 
up to $10^3$ K. 
This results from the expressions of the Jeans mass and the free fall
timescale, which imply that the average inflow
rate and the temperature of the collapsing cloud are increasing 
simultaneously.

There is a variety of results on the temperature of UC HII regions
(Churchwell \cite{Chu99} and Henning et al. \cite{Henn00}).
From their infrared emission, Churchwell finds, around them, a sizeable  ($\sim 10^{16}$ cm) dust evacuated cavity  and he estimates that the temperature at the inner 
face  of the dust shell
is typically 300 K. In the extreme 
case of W3, a massive star forming region,
there is even a hard X-ray emission, implying T up to 7  10$^{7}$ K
in the wind-shocked cavity surrounding its central UC HII region 
(Hofner and Churchwell, \cite{HoCh97}).
The  above results are likely to concern
environments that have  been altered 
by the presence of existing massive stars.
Indeed, for the first 
stars to be formed one need to consider the temperature of the cloud 
core just before  star formation begins.
 In regions like Orion, the 
so-called massive dense cores, which presumably are 
the birthsites of massive stars, are not as hot as the UC HII 
regions studied by Churchwell (\cite{Chu99}) and Henning et al. 
(\cite{Henn00}) indicate them to be. The temperature is more likely 
to be well below 100K (Caselli \& Myers \cite{Cas95}). Thus,  
the  temperature of the molecular cloud is not necessarily the only
parameter responsible for the enhanced accretion rates necessary to form
massive stars.

In massive star forming regions, like Orion, the velocity width of an
observed line is dominated by non-thermal, supersonic motions (e.g. 
Caselli \& Myers \cite{Cas95}). Thus, there is a significant contribution
of turbulent motions to the support of the clouds. In 
equation (\ref{acc_rate}) for the accretion rate, the thermal sound
speed should be replaced by the sum of a thermal and non--thermal
contribution. Turbulence takes a long time to be dissipated  and persists
during a significant part of the formation process. Caselli 
\& Myers (\cite{Cas95}) find a higher density  and pressure in massive
cores  leading to a fast mass infall. Thus, it may be that the turbulence
and the higher density of the ambient gas favour higher accretion
rates in massive star forming molecular clouds (we are indebted to 
Prof. F. Palla for this  very important remark).

\begin{figure}[tb]
  \resizebox{\hsize}{!}{\includegraphics{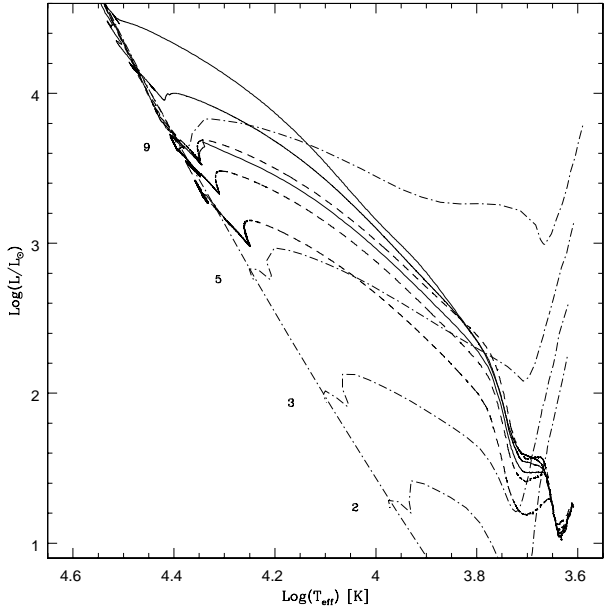}}
  \caption{ The continuous lines show the tracks calculated with expr. 4
for $\dot{M}_{\mathrm{ref}}$ = 10$^{-5}$ and 
for  $\varphi$ = 1.5, 1.0 and 0.5 from top to bottom. The broken lines
are the  birthlines calculated with expression 3 and F= 0.5, 1.0 and
1.5. The dot-broken lines are the PMS tracks for constant mass with the
indicated value (Bernasconi and Maeder \cite{BM96}).}
\label{puiscomp}
\end{figure}

There are  remarkable results concerning  the presence of huge,
likely bipolar, molecular outflows coming out of the regions of massive 
star formation. The luminosities of these regions are
estimated from the radio free-free fluxes and/or from the integrated
IR fluxes. The outflow rates come from the expansion velocities 
of the CO(1-0) line (Churchwell, \cite{Chu99}).
 The outflow rates $\dot{M}_{\rm{out}}$ behave continuously
like $L_{\rm{bol}}^{0.7}$ over 6 decades of 
luminosity (Shepherd and Churchwell, \cite{she96}).
Around the solar luminosity, the values of $\dot{M}_{\rm{out}}$ 
are about $10^{-5} M_{\odot}$ yr$^{-1}$, i.e. of the same order as the
currently estimated accretion rates. However for massive stars with
L from $10^4$ to $10^6 L_{\odot}$, the outflow
rates  $\dot{M}_{\rm{out}}$ are in the range of 10$^{-3}$ to 
$10^{-2} M_{\odot}$ yr$^{-1}$. There are several possible origins for these 
large outflows, however the review of the arguments by Churchwell 
(\cite{Chu99})
favours the possibility that the massive outflows are driven
by accretion, although it is not yet proved. From the large masses
present in the outflows and the luminosity of the central object,
he estimates that the fraction  $f$ of the infalling material 
incorporated into
the star is about 15 \%, while 85\% are deflected in the outflows.
Adopting a mass-luminosity relation of the form
 $L \sim M^3$, Churchwell (\cite{Chu99})
suggests that the outflow rates behave like
 $\dot{M}_{\rm{out}} \sim M^{2.1}$.
If we specifically consider, for example, the mass-luminosity
relation from the models by Schaller et al. (\cite{schaller})
in the broad mass interval
of 2 to 85 M$_{\odot}$, we would get

\begin{eqnarray}
\dot{M}_{accr} = 1.5 \; 10^{-5} f \left( \frac{M}{M_{\odot}} \right)^{1.54} \quad
\mathrm{M_{\odot} yr}^{-1} \; ,
\end{eqnarray}

\noindent
where $f$ is the accreted fraction of the infalling material.
Such results suggests that constant accretion rates of the
order of $10^{-5} M_{\odot}$ yr$^{-1}$ do not necessarily
 apply for all ranges
of stellar masses  and that much larger values of $\dot{M}_{\rm{accr}}$
may have to be considered for larger masses. If this is true,
several  of the arguments against the accretion
scenario for massive stars may not apply.

We also note  that there is a class of 
theoretical models of cloud equilibria which do in fact predict 
a strong dependence of the mass accretion rate on protostellar mass. 
These are the so-called logatropic spheres studied by 
McLaughlin \& Pudritz (\cite{McL96},\cite{McL97}) and Galli et al. 
(\cite{gal99}), where internal pressure varies like the
logarithm of the density. The equation of state departs from
isothermal, and the sound speed increases with decreasing density.
These models are useful to account for 
the line width density relation observed for molecular
clouds.  The accretion rate  onto a protostar is not
constant in logatropic models, but grows like $t^3$,
which implies that   $\dot{M}_{\rm{accr}}$ varies like
$M^{\frac{3}{4}}$  (McLaughlin \& Pudritz \cite{McL97}).
These models have also been applied to  study the collapse
of hot molecular cores leading to the formation of massive 
stars (Osorio et al. \cite{oso99}).

 In view of all the above arguments in favour of possible
large accretion rates,  we do think it is  worth to further
examine  the accretion
scenario for the formation of massive stars.

\section{Simple birthlines with constant or slowly varying accretion}

In this Section, we briefly show  for the purpose of comparison
some new sets of birthlines obtained with constant and slowly varying 
$\dot{M}_{\rm{accr}}$. The initial quasi-static model with
a mass $M_{\mathrm{ini}}$ = 0.7 M$_{\odot}$ are fully
convective (see Stahler et al. \cite{Sta88}; Bernasconi and Maeder, 
\cite{BM96}) and are started before the deuterium-burning sequence,
which is  treated with a time dependent convection scheme. It is well
known that the initial D-abundance and that of the accreting matter
influence the PMS evolution; here the initial and cosmic D abundance
is taken $5 \, \cdot \, 10^{-5}$ in mass fraction  
consistently with the results by Geiss (\cite{G93}). From many
test models by Bernasconi and Maeder (\cite{BM96}), it appears that the
exact value chosen for the initial model has no significant 
influence for the PMS structure and on the evolution of intermediate 
and massive stars. Only the age is possibly influenced by this choice
(see comments made about Tables 1, 2 and 3 below).

\subsection{Constant values of $\dot{M}_{\rm{accr}}$ }

Our simplest accretion model considers a constant mass accretion rate. In a recent study of the Orion Nebula, Palla and Stahler
(\cite{PS99}) have used constant values $\dot{M}_{\rm{accr}}$ 
equal to  $10^{-5} M_{\odot}$ yr$^{-1}$.
In Fig.1, the birthlines corresponding to $\dot{M}_{\rm{accr}}$
to 10$^{-6}$, 10$^{-5}$ and $10^{-4} M_{\odot}$ yr$^{-1}$ are
shown. Higher accretion rates lead to birthlines with higher luminosities.
 For high accretion rates,
a star gains more mass and thus luminosity during its PMS
contraction and reaches the ZAMS at a higher luminosity. Deuterium
is also continuously brought to the star and contributes to the
stellar luminosity (cf. Bernasconi and Maeder, \cite{BM96}).
Tout et al. (\cite{Tout99}) have recently shown that the PMS tracks
and age estimates of PMS stars are very much influenced by the accretion
rates and this is indeed in agreement with the present results.
\subsection{Slowly varying  $\dot{M}_{\rm{accr}}$}

\begin{table*}
\scriptsize
\caption{Birthline properties for accretion rate 
given by expr.4 with  $\varphi$ = 0.5  and  $\dot{M}_{\mathrm{ref}}$
= $10^{-5}$  M$_{\odot}$ yr$^{-1}$.
The columns are  in general self explicit; however, M$_\mathrm{core}$ gives the convective core extension (in mass fraction),
T$_{\rm{b}}$ is the temperature at the base of the 
convective envelope, when present; the surface content $^\mathrm{2}$H$_\mathrm{surf}$ in deuterium and the central hydrogen content
$^\mathrm{1}$H$_\mathrm{cent}$ are both given in mass fraction.}
\begin{tabular}{ccrcccrcccc}
\hline
& & & & & & & & & & \\
NB & Age & Mass & Log L & Log T$_\mathrm{e}$ & dM/dt  & L$_\mathrm{grav}$ & M$_\mathrm{core}$ & Log T$_\mathrm{b}$ & $^\mathrm{2}$H$_\mathrm{surf}$ & $^\mathrm{1}$H$_\mathrm{cent}$ \\
& yr & M$_{\odot}$ & L$_{\odot}$ & K & M$_{\odot}$ yr$^\mathrm{-1}$ & L$_{\odot}$ & M/M$_{\odot}$ & K &  &  \\
& & & & & & & & & & \\
\hline
& & & & & & & & & & \\
 1 & 0.000E+00 &  0.700 & 1.241 & 3.609 &  0.836E-06 &      17.36 &  1.000 & 5.832 & 5.00E-05  &  6.800E-01 \\
 2 & 3.271E+05 &  1.000 & 0.838 & 3.649 &  0.999E-06 &       4.70 &  1.000 & 6.250 & 5.68E-08  &  6.800E-01 \\
 3 & 1.156E+06 &  2.001 & 0.648 & 3.705 &  0.141E-05 &       1.22 &  0.000 & 6.474 & 4.87E-10  &  6.800E-01 \\
 4 & 1.791E+06 &  3.000 & 1.767 & 3.857 &  0.173E-05 &      55.02 &  0.000 & 
   & 4.20E-05  &  6.800E-01 \\
 5 & 2.327E+06 &  4.000 & 2.456 & 4.176 &  0.200E-05 &     133.86 &  0.132 &  
   & 1.10E-05  &  6.793E-01 \\
 6 & 2.799E+06 &  5.000 & 2.738 & 4.241 &  0.223E-05 &    -116.69 &  0.267 &  
   & 1.14E-05  &  6.780E-01 \\
 7 & 3.230E+06 &  6.011 & 3.077 & 4.294 &  0.244E-05 &    -145.66 &  0.249 & 
   & 5.00E-05  &  6.755E-01 \\
 8 & 3.628E+06 &  7.024 & 3.336 & 4.334 &  0.264E-05 &    -155.52 &  0.256 & 
   & 5.00E-05  &  6.730E-01 \\
 9 & 3.993E+06 &  8.022 & 3.489 & 4.362 &  0.282E-05 &    -171.74 &  0.291 & 
   & 5.00E-05  &  6.706E-01 \\
10 & 4.341E+06 &  9.037 & 3.646 & 4.388 &  0.299E-05 &    -199.69 &  0.319 & 
   & 5.00E-05  &  6.677E-01 \\
11 & 4.656E+06 & 10.008 & 3.790 & 4.411 &  0.315E-05 &    -219.64 &  0.335 & 
   & 5.00E-05  &  6.640E-01 \\
12 & 5.301E+06 & 12.147 & 4.055 & 4.450 &  0.346E-05 &    -263.28 &  0.331 & 
   & 5.00E-05  &  6.538E-01 \\
13 & 5.828E+06 & 14.047 & 4.247 & 4.478 &  0.372E-05 &    -301.07 &  0.356 & 
   & 5.00E-05  &  6.426E-01 \\
14 & 6.355E+06 & 16.085 & 4.425 & 4.502 &  0.398E-05 &    -366.45 &  0.385 & 
   & 5.00E-05  &  6.315E-01 \\
15 & 6.834E+06 & 18.058 & 4.572 & 4.522 &  0.422E-05 &    -426.49 &  0.411 & 
   & 5.00E-05  &  6.189E-01 \\
16 & 7.313E+06 & 20.146 & 4.708 & 4.539 &  0.446E-05 &    -486.96 &  0.470 & 
   & 5.00E-05  &  6.024E-01 \\
17 & 8.354E+06 & 25.080 & 4.974 & 4.568 &  0.499E-05 &    -646.14 &  0.465 & 
   & 5.00E-05  &  5.620E-01 \\
18 & 9.305E+06 & 30.062 & 5.189 & 4.585 &  0.547E-05 &    -915.03 &  0.486 & 
   & 5.00E-05  &  5.052E-01 \\
19 & 1.018E+07 & 35.073 & 5.375 & 4.595 &  0.591E-05 &   -1234.56 &  0.513 & 
   & 5.00E-05  &  4.564E-01 \\
20 & 1.099E+07 & 40.028 & 5.530 & 4.590 &  0.632E-05 &   -1702.94 &  0.530 & 
   & 5.00E-05  &  3.818E-01 \\
21 & 1.175E+07 & 45.018 & 5.670 & 4.569 &  0.670E-05 &   -2319.24 &  0.503 & 
   & 5.00E-05  &  2.939E-01 \\
22 & 1.248E+07 & 50.009 & 5.798 & 4.512 &  0.706E-05 &   -3345.30 &  0.467 & 
   & 5.00E-05  &  1.699E-01 \\
23 & 1.317E+07 & 55.012 & 5.923 & 4.395 &  0.741E-05 &   -2531.52 &  0.417 & 
   & 5.00E-05  &  2.512E-02 \\
& & & & & & & & & & \\
\hline
\hline
\end{tabular}
\end{table*}

\begin{table*}
\scriptsize
\caption{ Same as in Table 1, but for $\varphi$ = 1.0 .}
\begin{tabular}{ccrcccrcccc}
\hline
& & & & & & & & & & \\
NB & Age & Mass & Log L & Log T$_\mathrm{e}$ & dM/dt  & L$_\mathrm{grav}$ & M$_\mathrm{core}$ & Log T$_\mathrm{b}$ & $^\mathrm{2}$H$_\mathrm{surf}$ & $^\mathrm{1}$H$_\mathrm{cent}$ \\ \\
& yr & M$_{\odot}$ & L$_{\odot}$ & K & M$_{\odot}$ yr$^\mathrm{-1}$ & L$_{\odot}$ & M/M$_{\odot}$ & K & & \\                 
& & & & & & & & & & \\
\hline
& & & & & & & & & & \\
 1 & 0.000E+00 &  0.700 & 1.241 & 3.609 &  0.700E-06 &      17.36 &  1.000 & 5.832 & 5.00E-05  &  6.800E-01 \\
 2 & 3.575E+05 &  1.000 & 0.793 & 3.649 &  0.999E-06 &       4.03 &  1.000 & 6.274 & 2.80E-08  &  6.800E-01 \\
 3 & 1.050E+06 &  2.001 & 0.666 & 3.704 &  0.199E-05 &       0.05 &  0.000 & 6.427 & 1.24E-09  &  6.800E-01 \\
 4 & 1.456E+06 &  3.000 & 1.558 & 3.802 &  0.299E-05 &      29.24 &  0.000 & 
   & 4.79E-05  &  6.800E-01 \\
 5 & 1.744E+06 &  4.002 & 2.776 & 4.197 &  0.399E-05 &     -33.63 &  0.197 & 
   & 1.04E-05  &  6.799E-01 \\
 6 & 1.967E+06 &  5.000 & 2.661 & 4.213 &  0.499E-05 &      89.33 &  0.211 & 
   & 1.05E-05  &  6.793E-01 \\
 7 & 2.149E+06 &  6.000 & 3.086 & 4.305 &  0.599E-05 &    -374.60 &  0.293 & 
   & 1.08E-05  &  6.783E-01 \\
 8 & 2.308E+06 &  7.031 & 3.280 & 4.331 &  0.699E-05 &    -425.53 &  0.276 & 
   & 5.00E-05  &  6.770E-01 \\
 9 & 2.445E+06 &  8.059 & 3.515 & 4.366 &  0.798E-05 &    -516.27 &  0.266 & 
   & 5.00E-05  &  6.756E-01 \\
10 & 2.561E+06 &  9.046 & 3.680 & 4.393 &  0.896E-05 &    -564.90 &  0.278 & 
   & 5.00E-05  &  6.744E-01 \\
11 & 2.667E+06 & 10.056 & 3.800 & 4.415 &  0.996E-05 &    -667.72 &  0.313 & 
   & 5.00E-05  &  6.735E-01 \\
12 & 2.851E+06 & 12.073 & 4.034 & 4.452 &  0.119E-04 &    -880.94 &  0.359 & 
   & 5.00E-05  &  6.712E-01 \\
13 & 3.005E+06 & 14.082 & 4.236 & 4.483 &  0.139E-04 &   -1084.87 &  0.365 & 
   & 5.00E-05  &  6.682E-01 \\
14 & 3.141E+06 & 16.113 & 4.405 & 4.508 &  0.159E-04 &   -1159.10 &  0.412 & 
   & 5.00E-05  &  6.660E-01 \\
15 & 3.257E+06 & 18.085 & 4.546 & 4.528 &  0.179E-04 &   -1623.84 &  0.439 & 
   & 5.00E-05  &  6.635E-01 \\
16 & 3.363E+06 & 20.104 & 4.672 & 4.546 &  0.199E-04 &   -1725.76 &  0.466 & 
   & 5.00E-05  &  6.606E-01 \\
17 & 3.586E+06 & 25.084 & 4.921 & 4.579 &  0.248E-04 &   -2599.08 &  0.503 & 
   & 5.00E-05  &  6.545E-01 \\
18 & 3.769E+06 & 30.116 & 5.116 & 4.603 &  0.298E-04 &   -3324.35 &  0.566 & 
   & 5.00E-05  &  6.469E-01 \\
19 & 3.924E+06 & 35.128 & 5.272 & 4.621 &  0.347E-04 &   -4241.38 &  0.606 & 
   & 5.00E-05  &  6.408E-01 \\
20 & 4.059E+06 & 40.193 & 5.401 & 4.635 &  0.398E-04 &   -5319.48 &  0.640 & 
   & 5.00E-05  &  6.351E-01 \\
21 & 4.175E+06 & 45.113 & 5.509 & 4.645 &  0.446E-04 &   -6289.63 &  0.669 & 
   & 5.00E-05  &  6.293E-01 \\
22 & 4.278E+06 & 49.946 & 5.600 & 4.654 &  0.491E-04 &   -6833.45 &  0.656 & 
   & 5.00E-05  &  6.262E-01 \\
23 & 4.376E+06 & 55.062 & 5.682 & 4.660 &  0.546E-04 &   -8946.76 &  0.715 & 
   & 5.00E-05  &  6.194E-01 \\
24 & 4.466E+06 & 60.217 & 5.758 & 4.665 &  0.597E-04 &  -10319.23 &  0.686 & 
   & 5.00E-05  &  6.144E-01 \\
25 & 4.622E+06 & 70.340 & 5.887 & 4.671 &  0.698E-04 &  -6300.715 &  0.718 & 
   & 5.00E-05  &  6.097E-01 \\
26 & 4.753E+06 & 80.182 & 5.991 & 4.683 &  0.796E-04 &  -6455.416 &  0.739 & 
   & 5.00E-05  &  6.012E-01  \\
27 & 4.977E+06 & 100.23 & 6.160 & 4.689 &  0.996E-04 & -23297.530 &  0.768 & 
   & 5.00E-05  &  5.852E-01 \\
& & & & & & & & & & \\
\hline
\hline
\end{tabular}
\end{table*}

\begin{table*}
\scriptsize
\caption{ Same as in Table 1, but for $\varphi$ = 1.5 .}
\begin{tabular}{ccrcccrcccc}
\hline
& & & & & & & & & & \\
NB & Age & Mass & Log L & Log T$_\mathrm{e}$ & dM/dt  & L$_\mathrm{grav}$ & M$_\mathrm{core}$ & Log T$_\mathrm{b}$ & $^\mathrm{2}$H$_\mathrm{surf}$ & $^\mathrm{1}$H$_\mathrm{cent}$ \\
& yr & M$_{\odot}$ & L$_{\odot}$ & K & M$_{\odot}$ yr$^\mathrm{-1}$ & L$_{\odot}$ & M/M$_{\odot}$ & K & & \\                 
& & & & & & & & & & \\
\hline
& & & & & & & & & & \\
 1 & 0.000E+00 &  0.700 & 1.241 & 3.609 &  0.585E-06 &      17.36 &  1.000 & 5.832 & 5.00E-05  &  6.800E-01  \\ 
 2 & 3.907E+05 &  1.000 & 0.745 & 3.649 &  0.999E-06 &       3.42 &  1.000 & 6.297 & 1.46E-08  &  6.800E-01  \\
 3 & 9.767E+05 &  2.000 & 0.721 & 3.708 &  0.282E-05 &      -1.22 &  0.000 & 6.129 & 5.89E-06  &  6.800E-01  \\
 4 & 1.236E+06 &  3.000 & 1.120 & 3.766 &  0.519E-05 &       7.92 &  0.000 & 6.065 & 1.68E-05  &  6.800E-01  \\
 5 & 1.391E+06 &  4.002 & 2.548 & 4.067 &  0.799E-05 &     340.19 &  0.000 & 
   & 2.11E-05  &  6.800E-01  \\
 6 & 1.496E+06 &  5.000 & 3.018 & 4.268 &  0.111E-04 &    -476.96 &  0.350 & 
   & 1.13E-05  &  6.797E-01  \\
 7 & 1.574E+06 &  6.000 & 2.944 & 4.266 &  0.146E-04 &   -1851.30 &  0.454 &
   & 1.12E-05  &  6.794E-01  \\
 8 & 1.635E+06 &  7.001 & 3.297 & 4.342 &  0.185E-04 &   -1275.61 &  0.340 & 
   & 1.17E-05  &  6.790E-01  \\
 9 & 1.684E+06 &  8.004 & 3.560 & 4.369 &  0.226E-04 &   -1535.61 &  0.263 & 
   & 5.00E-05  &  6.784E-01  \\
10 & 1.725E+06 &  9.038 & 3.774 & 4.402 &  0.269E-04 &   -1468.14 &  0.255 & 
   & 5.00E-05  &  6.778E-01  \\
11 & 1.760E+06 & 10.053 & 3.829 & 4.419 &  0.311E-04 &   -2105.96 &  0.288 & 
   & 5.00E-05  &  6.773E-01  \\
12 & 1.818E+06 & 12.164 & 4.065 & 4.457 &  0.413E-04 &   -3042.56 &  0.335 & 
   & 5.00E-05  &  6.766E-01  \\
13 & 1.857E+06 & 13.961 & 4.220 & 4.484 &  0.508E-04 &   -3639.83 &  0.359 & 
   & 5.00E-05  &  6.760E-01  \\
14 & 1.896E+06 & 16.185 & 4.395 & 4.511 &  0.632E-04 &   -5075.49 &  0.423 & 
   & 5.00E-05  &  6.757E-01  \\
15 & 1.925E+06 & 18.220 & 4.538 & 4.532 &  0.754E-04 &   -6693.96 &  0.460 & 
   & 5.00E-05  &  6.751E-01  \\
16 & 1.949E+06 & 20.222 & 4.662 & 4.550 &  0.880E-04 &   -8078.64 &  0.484 & 
   & 5.00E-05  &  6.746E-01  \\
17 & 1.997E+06 & 25.350 & 4.915 & 4.585 &  0.123E-03 &  -12403.20 &  0.535 & 
   & 5.00E-05  &  6.732E-01  \\
18 & 2.031E+06 & 30.187 & 5.099 & 4.609 &  0.159E-03 &  -17836.61 &  0.577 & 
   & 5.00E-05  &  6.721E-01  \\
19 & 2.060E+06 & 35.516 & 5.261 & 4.629 &  0.203E-03 &  -24182.66 &  0.616 & 
   & 5.00E-05  &  6.711E-01  \\
20 & 2.079E+06 & 39.887 & 5.372 & 4.642 &  0.241E-03 &  -30703.26 &  0.644 & 
   & 5.00E-05  &  6.704E-01  \\
21 & 2.099E+06 & 45.110 & 5.485 & 4.655 &  0.289E-03 &  -39344.84 &  0.672 & 
   & 5.00E-05  &  6.697E-01  \\
22 & 2.113E+06 & 49.724 & 5.571 & 4.664 &  0.333E-03 &  -47801.27 &  0.697 & 
   & 5.00E-05  &  6.692E-01  \\
23 & 2.128E+06 & 55.077 & 5.658 & 4.673 &  0.388E-03 &  -58971.57 &  0.718 & 
   & 5.00E-05  &  6.685E-01  \\
24 & 2.142E+06 & 61.335 & 5.746 & 4.681 &  0.454E-03 &  -72931.53 &  0.737 & 
   & 5.00E-05  &  6.679E-01  \\
25 & 2.161E+06 & 71.464 & 5.869 & 4.691 &  0.569E-03 &  -98576.05 &  0.767 & 
   & 5.00E-05  &  6.670E-01  \\
26 & 2.176E+06 & 80.783 & 5.964 & 4.699 &  0.682E-03 & $<$ -1.0E+05 &  0.796 & 
   & 5.00E-05  &  6.664E-01 \\
27 & 2.200E+06 & 100.86 & 6.128 & 4.710 &  0.945E-03 &  $<$ -1.0E+05 &  0.833 & 
   & 5.00E-05  &  6.653E-01 \\
& & & & & & & & & & \\
\hline
\hline
\end{tabular}
\end{table*}

The models by Bernasconi and Maeder (\cite{BM96}) did not use
constant accretion rates, but the values of 
$\dot{M}_{\rm{accr}}$ were slightly increasing with the already
accreted mass, according to the prescriptions of a simple model
of collapsing clouds. This model needs several input parameters
such as the temperature T of the cloud (currently 30 K), the  mean molecular
weight $\mu$ (currently 2.4). The effect of turbulent pressure were accounted for according to the velocity dispersion vs. size of the clouds
given by Larson (\cite{larson}). In Fig. 1, a model
with these prescriptions  is  also shown (model with F=1.0).
We also show two other models with rates 10 times smaller and 10
times larger than the rates  
$\dot{M}_{\rm{BM}}$  by  Bernasconi and Maeder (\cite{BM96}), according
to the expression 

\begin{eqnarray}
\dot{M}_{\rm{accr}} =  F  \: \dot{M}_{\rm{BM}}
\end{eqnarray}

Fig. 1 indicates clearly the differences  between the two sets of models.
These differences  are
small, especially for the low accretion rates.
The model with F=0.1 reaches the ZAMS at about 4.4 M$_{\odot}$, 
the one with F=1.0 at 9.5  M$_{\odot}$, and at 27.5  M$_{\odot}$ for F = 10.0. After having reached the ZAMS, the continuously accreting protostar carries on its evolution along the ZAMS.

\begin{figure*}[tb]
  \resizebox{14cm}{!}{\includegraphics{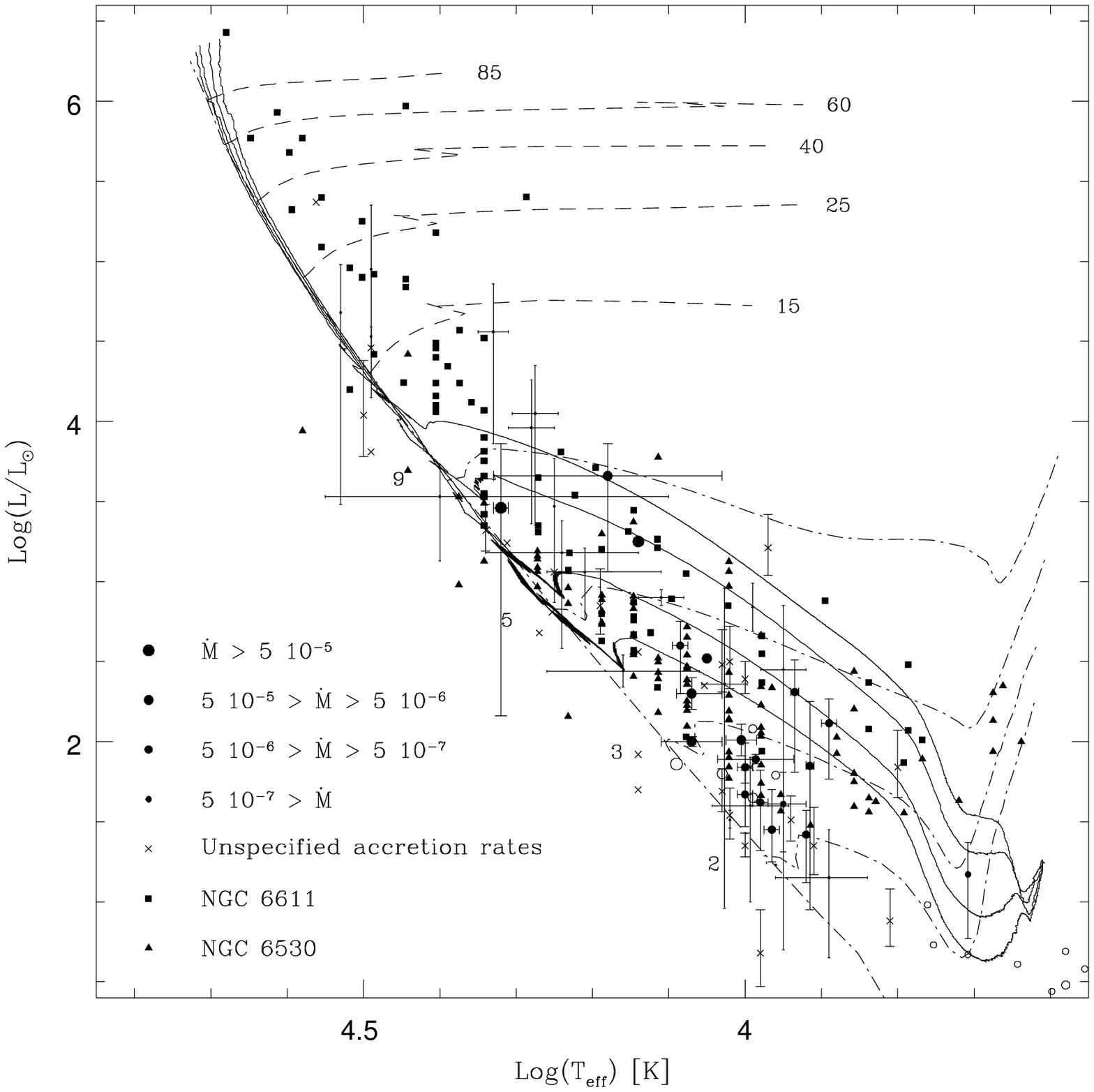}}
  \caption{ The continuous lines represent birthlines obtained with $\varphi$
= 1.0 and $\dot{M}_{\rm{ref}}$ being equal to 10$^{-6}$, 2 10$^{-6}$, 5 $\cdot$
10$^{-6}$ and
 10$^{-5}$ M$_{\odot}$ yr$^{-1}$ from bottom to top 
respectively. The higher $\dot{M}_{\rm{ref}}$ is the 
higher the birthline is in the HR diagram. The other curves have the same meaning as in Fig. 2 
and the observations are the same.  }
  \label{exp1}
\end{figure*}

\begin{figure*}[tb]
  \resizebox{14cm}{!}{\includegraphics{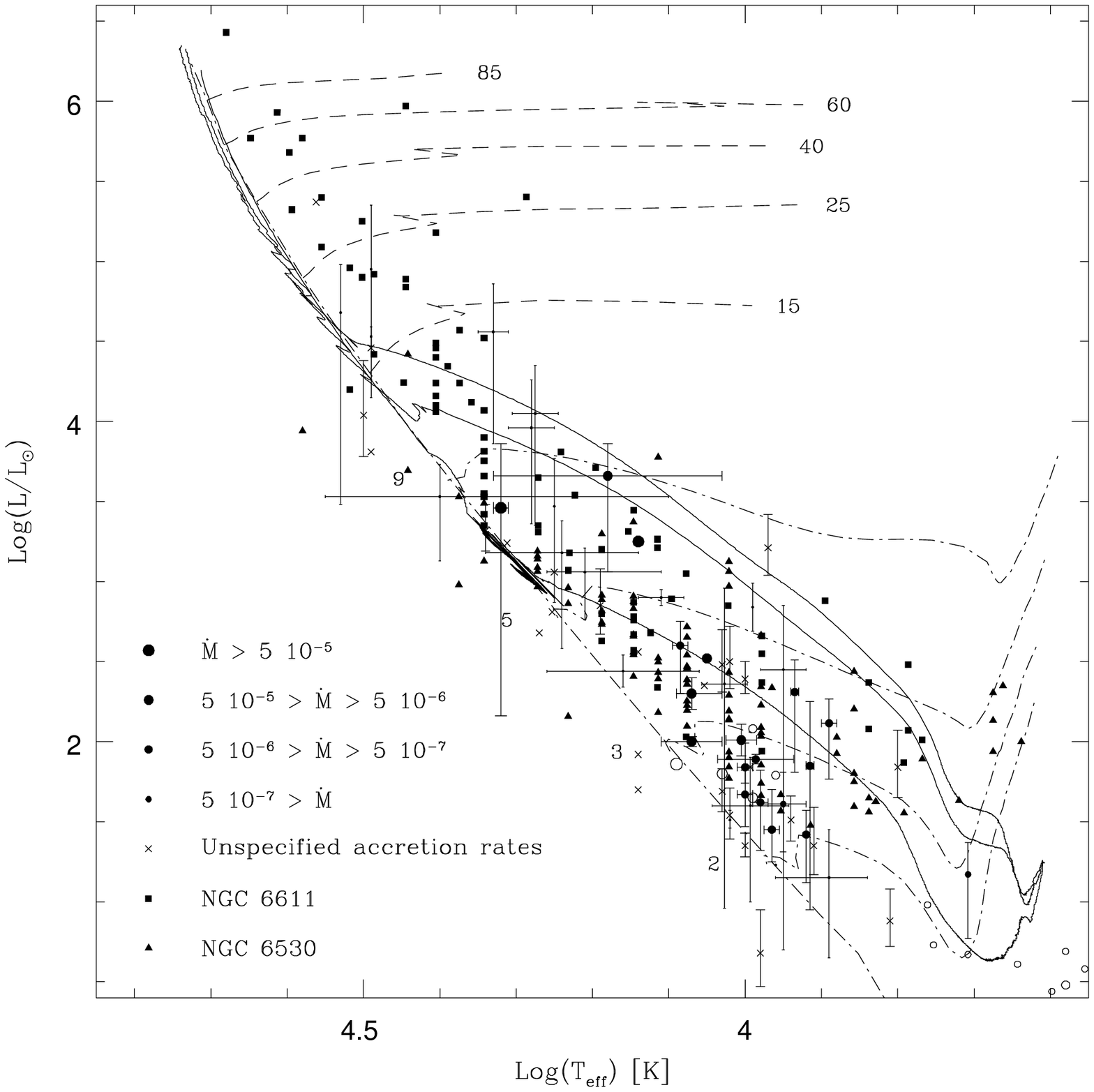}}
  \caption{The continuous lines represent birthlines obtained with $\varphi$
= 1.5 and $\dot{M}_{\rm{ref}}$ being equal to 10$^{-6}$, 5 $\cdot$ 10$^{-6}$ and
 10$^{-5}$ M$_{\odot}$ yr$^{-1}$ respectively. Same remarks as for Fig.4 }
  \label{exp1.5}
\end{figure*}

On the upper  main sequence, the tracks are also influenced by the
accretion rates. For low $\dot{M}_{\rm{accr}}$, the time necessary to
build up
a massive star is so long that the star begins to burn a large
fraction of its central hydrogen. Thus it has already moved away 
from the ZAMS, when it becomes visible at the end of its very long 
accretion phase. For high accretion rates, the time 
necessary to form  a massive star is so short that a small amount of hydrogen is burnt
and therefore the star becomes visible close to the ZAMS. Theses features
are well shown in Fig. 1.

In Fig. 2, we compare 12 birthlines made with F values between 0.1 and 5.0
with recent observations of PMS stars in various clusters (the references
are given in the figure caption). Indications are given for stars which have observed values of the accretion rates.  Let us
note with caution that there are considerable uncertainties in the derivations
of luminosities and T$_{\rm{eff}}$, as shown by the error bars. The birthlines
have to be seen as upper envelopes, since a fraction of the stars
may already have ended their accretion phase and are moving towards the
ZAMS along the canonical PMS tracks. We do not know whether one single birthline should apply to
various star groups; however let us for simplicity adopt such
a  view.

From Fig.2, we can make the following remarks. --1. A birthline with 
$F$ between 1 and 5 fits best as an upper envelope, this corresponding
to $\dot{M}_{\rm{accr}}$ in the range of 
 10$^{-5}$ and $10^{-4} M_{\odot}$ yr$^{-1}$. Thus, it is very 
likely that accretion rates  higher
than the currently used values of  $10^{-5} M_{\odot}$ yr$^{-1}$ are
necessary, at least for the most luminous PMS stars.
 --2.  The upper part of the tracks at the time the star becomes
visible also depends on the previous accretion rates (Figs. 1 and 3).
There is some support towards high $\dot{M}_{\rm{accr}}$
from the observations by Hanson et al. (\cite{hanson}). 
They find some very massive PMS objects
(M $\geq$ 60 M$_{\odot}$)
close to the ZAMS and as seen above this also constrains the
accretion rates. Thus these stars have to be formed in a time of the order of
10$^6$ yr  at most. This implies an \textit{average} accretion rate
of about 10$^{-4}$ M$_{\odot}$ yr$^{-1}$. 
Thus,  a most critical 
difficulty for all these  birthlines with slowly increasing
 $\dot{M}_{\rm{accr}}$ is their too long lifetimes 
(cf. also Hanson et al. \cite{hanson}). --3. The 
observations suggest that the birthline should join the ZAMS 
between about 9 and 15 M$_{\odot}$.
--4. The distribution of stars in the HR diagram do not tightly constrain
 the slope of the birthline. However, we note that 
the observed distribution of stars in Fig. 2 seems to suggest a slope,
 which may be steeper 
than the one predicted by Bernasconi and  Maeder (\cite{BM96}). Especially
 the birthline could be lower at lower luminosities.

The various difficulties met with these accretion models as a formation
mechanism of massive stars lead us to examine in the next Section models 
where the accretion rate $\dot{M}_{\rm{accr}}$ increases more rapidly
with mass. Indeed this will give rise to birthlines with steeper
slopes in the HR diagram and at the same time shorten the formation timescales.

\section{Models with strongly increasing accretion rates}

The results of the two previous Sections suggest that the accretion rates
may increase relatively quickly with the stellar mass. To further test this
hypothesis, we examine the consequences of a power law with exponent $\varphi$

\begin{eqnarray}
\label{puis}
\dot{M}_{\rm{accr}}(M)  =  \dot{M}_{\rm{ref}}  \left( \frac{M}{M_{\odot}} \right)^{\varphi}
\end{eqnarray}

\noindent
 We consider that the accretion rate is increasing with an exponent
 $\varphi$ of the mass. Models with
$\varphi$ = 0.5, 1.0 and 1.5 are calculated. We provide in the
Tables 1, 2 and 3 some important data for these models, assuming a
reference mass loss rate $\dot{M}_{\rm{ref}}$ of 10$^{-5}$ M$_{\odot}$
yr$^{-1}$. We shall also test this value below. The models are started
with an initial mass $M_{\mathrm{ini}}$ = 0.7 M$_{\odot}$ and the zero
of the age scales are placed at this time. Another possible choice
(Bernasconi, \cite{Ber96}) would be to take as initial
age $\frac{0.7 M_{\odot}}{<\dot{M}_{\rm{accr}}>}$,
where $<\dot{M}_{\rm{accr}}>$ represents the average accretion
rate since the start of the formation process. As a matter of fact, we
 do not strictly know the zero point in the age scale, and therefore
 any convention is slightly arbitrary.

Fig. 3 shows 3 birthlines obtained with expression \ref{puis}. We notice that
the tracks by Bernasconi and Maeder (\cite{BM96}; corresponding to  F = 1.0)
are rather close to those with $\varphi$ = 0.5. For higher
values of $\varphi$, we obtain, as expected, a much steeper slope
of the birthlines. 
As shown in the Tables 1, 2 and 3, the models with a higher 
$\varphi$ have much shorter formation
times. In models of high $\varphi$
the central H-content is still much higher,  and the size of their
convective cores are larger. Both points  are consistent
with  the fact that the  ages are shorter
 and the evolution much less advanced.
For $\varphi$ = 0.5, and even more for the constant $\dot{M}_{\rm{accr}}$, 
the PMS lifetimes t$_\mathrm{PMS}$ are longer than 
the Main Sequence (MS) lifetimes t$_\mathrm{MS}$ 
for massive stars above about 20 M$_{\odot}$. In this case,
 t$_\mathrm{MS}$  = 8.1  10$^6$ yr, and t$_\mathrm{PMS}$   = 7.3   10$^6 yr$.
This is a  very severe problem for slowly increasing  $\dot{M}_{\rm{accr}}$
and a very  important argument in favour of models with an
accretion rate  quickly increasing with stellar mass.
For $\varphi$ = 1.0, the equality of the two lifetimes 
t$_\mathrm{PMS}$   and  t$_\mathrm{MS}$  is realized around 
a mass of about 45 M$_{\odot}$; however a large
fraction of t$_\mathrm{PMS}$  is still spent in the low mass regime.

For $\varphi$ = 1.5, the situation with respect
to the lifetimes is  much more satisfactory, as t$_\mathrm{PMS}$  is shorter
than t$_\mathrm{MS}$  up to at least 120 M$_{\odot}$.  Moreover, half
of t$_\mathrm{PMS}$   is spent below a mass of 2 M$_{\odot}$, and the time for
the star to evolve 
 from 2 to 120 M$_{\odot}$ is  about 10$^6$ yr, which is an
acceptable value. Due to this we must stress
that accretion rates strongly increasing with mass become
 a necessity in order to form stars on a timescale significantly
smaller than the MS lifetime, so that they have not evolved too
far from the ZAMS when they become visible. Also, a too long lifetime
will allow a massive star to ionize a too large surrounding
region, preventing any further accretion. In this respect, we emphasize that
the models with
$\varphi \simeq$ 1.5  are much more favourable than the other ones
studied here.

These calculations raise
the question whether, at a given value of the stellar mass,
 the  initial formation of a massive star 
occurs with the same accretion rate as for a star with a
low final mass. This is what is assumed here.
 This looks 
reasonable, since according to Newton's Theorem the gravitational potential
in a spherical configuration is determined by the matter within the considered radius. However, further studies may
show the importance of environmental effects, such as the local
density and temperature in a cluster, which are not taken into account 
in the present study.

In order to further examine the accretion rates and their dependence
on $\varphi$ and $\dot{M}_{\rm{ref}}$
in expr. 4, some
other models have been calculated. Fig. 4 shows models for $\varphi$
= 1.0 and $\dot{M}_{\rm{ref}}$ =
 10$^{-6}$, 2 $\cdot$ 10$^{-6}$, 5 $\cdot$ 10$^{-6}$ and
 10$^{-5}$  M$_{\odot}$ yr$^{-1}$ respectively. These birthlines are compared to the observations already shown in Fig. 2.
Clearly the highest value for $\dot{M}_{\rm{ref}}$  
gives the best fit as an upper envelope.
 This confirms that
in the range of  intermediate mass stars the accretion rates
are not as low as 10$^{-5}$ M$_{\odot}$ yr$^{-1}$.
As already noticed for Fig. 2, it is difficult to obtain constraints
on the slope $\varphi$ from the point distributions in the HR diagram.
However, the agreement for the highest curve in Fig. 4
is rather better than for Fig.2. Also, it is possible that
 the theoretical slope 
is not steep enough in view of the observations. 
 From the envelopes in the HR diagram, the values
of $\dot{M}_{\rm{ref}}$ are much better determined  than the value
of $\varphi$.

 Fig. 5 shows the same 
kind of results, but for an exponent 
$\varphi$ = 1.5 and $\dot{M}_{\rm{ref}}$ =  
10$^{-6}$, 5 $\cdot$ 10$^{-6}$ and 10$^{-5}$  M$_{\odot}$ yr$^{-1}$ respectively. Clearly the two upper curves
give the best fit,  and maybe the highest one is the best. 
From the  envelope fits in the HR diagram, it is hard to say
 whether a value  $\varphi$ = 1.0 or 1.5 is best, however from
the considerations on the lifetimes, the case with $\varphi$ = 1.5
is clearly favoured.  Therefore, our preferred choice of parameters
for the model given by expr. 4 is an exponent $\varphi \simeq $ 1.5
and a multiplying factor $\dot{M}_{\rm{ref}} \simeq $ 10$^{-5}$  M$_{\odot}$ yr$^{-1}$.

 We notice 
 that it is amazing how this slope and the multiplying 
factor are close to the results obtained by Churchwell (\cite{Chu99})
and Henning et al. (\cite{Henn00}) and in particular to the
values given in expr. 2.  Without saying that the  physical
behaviour of the accretion rates is  determined by a law
which is exactly of the form given by expr. 4, it is interesting that these
two very different approaches give a rather similar dependence with
 respect to the stellar mass. Further theoretical and 
observational analyses are needed to give more insight into the dependence
of the accretion rate on various possible parameters.

\section{Conclusions and remarks on the maximum stellar mass}

The main result of this work is that the accretion rate of a 
forming protostar strongly depends on the protostar's mass.
A birthline described by a power law 

\begin{equation}
\dot{M}_{\rm{accr}} = \dot{M}_{\rm{ref}} \left(\frac{M}{M_{\odot}}
\right)^{\varphi}
\end{equation}

\noindent
with  $\dot{M}_{\rm{ref}} \simeq $ 10$^{-5}$ and 
$\varphi$ = 1.5 gives the best upper envelope
for PMS stars in the HR diagram and, more importantly, 
it satisfies also the constraints coming from the formation 
lifetimes. The above law is also quite consistent  
with radio and IR  studies of protostellar outflows 
(Churchwell, \cite{Chu99};
Henning et al., \cite{Henn00}), which show $\dot{M}_{\rm{accr}}$ 
quickly increasing with the luminosity of the UC HII region.

It is interesting that the high values of $\dot{M}_{\rm{accr}}$ 
suggested here well correspond to the permitted domain of 
accretion rates found by Wolfire and Cassinelli (\cite{WC87}).
The limits of this permitted domain are defined, on the low side of
the values 
of $\dot{M}_{\rm{accr}}$, by the condition that the momentum
in the accretion flow is larger than the outwards radiation momentum.
On the high side, it is fixed by the condition that the shock 
luminosity due to the accretion process is smaller than the Eddington
luminosity. In addition, Wolfire and Cassinelli  (\cite{WC87})
also found, for the occurrence of inflows onto massive stars, that
the dust abundance has to be  reduced by at least
a factor of 4 and that the larger graphite grains are absent
from the dust distribution function. In the context of the star models
presented here, which do not follow the properties of the 
surrounding interstellar matter, we do not know whether this 
additional condition is met.

If this accretion scenario for forming massive stars proves to 
be the correct one, it has several further implications: on the 
luminosities and T$_{\rm{eff}}$ of the progenitors of massive stars;
on the lifetimes
of PMS evolution; on the initial stellar structure on the ZAMS;
on surface abundances of light elements; etc... It has also an
impact on the slope of the initial mass function (IMF), since the final
mass spectrum is not only shaped by the size of the collapsing
fragments, which determine the reservoir of matter  potentially
available, but also by the accretion process which leads the star
to reach its final mass. 

Moreover the maximum stellar mass is determined by the
 physics intervening
in the accretion process. In particular,
 the maximum stellar mass is the mass for which the accretion rate
$\dot{M}_{\rm{accr}}$ is such that the shock luminosity L$_{\rm{shock}}$
due to the accretion plus the protostellar luminosity L$_{*}$ 
is equal to the Eddington luminosity L$_{\mathrm{Edd}}$, i.e.
L$_{\mathrm{shock}}$ + L$_{*}$ = L$_{\mathrm{Edd}}$. 
If R$_{\mathrm{shock}}$ is the radius where the 
shock occurs, one has: 

\begin{eqnarray}
\frac{G \dot{M}_{\mathrm{accr}} M}{R_{\mathrm{shock}}} + L_{*} = 
\frac{4 \pi c G M}{\kappa_{\mathrm{dust}}}.
\end{eqnarray}

\noindent
The relevant opacity to be considered here is the dust opacity
$\kappa_{\mathrm{dust}}$
near the inner face of the dust cavity, since the grain opacity is the 
largest opacity source which may prevent further material accretion.
As shown by Pollack et al. (\cite{Poll94}), the main opacity source 
are organics  below the vaporization temperature (T $\simeq $ 500 K)
and silicates  and metallic iron at higher temperatures.
The opacity to be considered is likely the Planck opacity, i.e. the
flux weighted opacity appropriate for small optical depths.
The typical values of the Planck opacity  $\kappa_{\mathrm{dust}}$ 
range between 2 and 8 cm$^2$/g.
Now, for the high accretion rates considered in the upper mass range,
 the shock luminosity dominates over the stellar luminosity 
by about 3 to 4 orders of magnitudes so that we may ignore L$_{*}$. 
Thus, we obtain a simple expression for the limiting accretion rate

\begin{eqnarray}
\dot{M}_{\mathrm{accr}}  = 
\frac{4 \pi c \; R_{\mathrm{shock}}}{\kappa}.
\end{eqnarray}

\noindent
If we consider for simplicity that the shock radius is some multiple
$\alpha$ of the stellar radius, we can apply the
mass-radius relation obtained from the models of Schaller et
al. (\cite{schaller}). The relation, valid between 40 and
120 M$_{\odot}$, is thus:

\begin{eqnarray}
\frac{R_\mathrm{shock}}{R_{\odot}} =  \alpha \;
\left( \frac{M}{M_{\odot}} \right)^{0.557}
\end{eqnarray}

\noindent
Now, we search for the intersection of the expression (5) for 
$\dot{M}_{\mathrm{accr}}(M)$ with relation  (7) also accounting
for (8). This gives the maximum possible stellar mass

\begin{equation}
\frac{M_\mathrm{shock}}{M_{\odot}} = 
\left(\frac{41.6 \; \alpha}{\kappa_\mathrm{dust}} \right)^{1.06}
\end{equation}

\noindent
with $\kappa_\mathrm{dust}$ expressed in cm$^2$/g.
The main interest is to emphasize that in the present context the
maximum mass is defined by the intersection of
$\dot{M}_{\mathrm{accr}}(M)$ with the the accretion rate (7)
giving a shock luminosity equal to the appropriate Eddington luminosity.
More detailed models of the stellar surroundings are of course necessary.
The above simple derivation  may at most give an order of 
magnitude, if we know both the location of R$_\mathrm{shock}$ given by
$\alpha$ and $\kappa_\mathrm{dust}$. While the approximate 
range of values of 
$\kappa_\mathrm{dust}$ is known as seen above, the value of
$\alpha$ is  uncertain, since it depends on 
the adopted structure and turbulence of the clouds (cf. Pollack et al.,
\cite{Poll94}). As an example with $\alpha \simeq 10$,
we would have a maximum mass between 70 and 300  M$_{\odot}$.

The accretion scenario  thus leads to a new simple concept for
the maximum stellar mass. A change of metallicity will influence
both the opacity and the location of the shock radius, thus 
the resulting effect of metallicity cannot be estimated without detailed
models of collapsing clouds.

\begin{acknowledgements} The authors  want to express their
 gratitude to Prof. F. Palla for his most useful remarks on the manuscript.
They also express their thanks
to Dr. Georges Meynet for his advice and helpful discussions. \end{acknowledgements}


\begin{thebibliography}{}
\bibitem[1998]{anck98}
van den Ancker, M.E., de Winter, D., Tjin A Djie, H.R.E., 1998, A\&A 330, 145
\bibitem[1997a]{anck97a}
van den Ancker, M.E., Th\'e, P.S., Tjin A Djie, H.R.E., et al., 1997a, A\&A 324, L33
\bibitem[1997b]{anck97b}
van den Ancker, M.E., Th\'e, P.S., Feinstein, A., et al., 1997b, A\&AS 123, 63
\bibitem[1994]{BeeM94}
Beech, M, Mitalas, R., 1994, ApJS 95, 517
\bibitem[1996]{Ber96}
Bernasconi, P.A., 1996, A\&AS 120, 57
\bibitem[1996]{BM96}
Bernasconi, P.A., Maeder, A., 1996, A\&A 307, 829
\bibitem[1992]{ber92}
Berrilli, F., Corcuilo, G., Ingrosso, G., et al., 1992, ApJ 398, 254
\bibitem[1998]{Bo98}
Bonnell, I.A., Bate, M.R., Zinnecker, H., 1998, MNRAS 298, 93
\bibitem[1995]{Cas95}
Caselli, P., Myers, P.C., 1995, ApJ 446, 665
\bibitem[1999]{Chu99}
Churchwell, E., 1999, "Massive Star Formation: The Role of Bipolar Outflows'', in UN solved Problems in Stellar Evolution, Proc. STScI Meeting, Ed. M. Livio, in press
\bibitem[1979]{coh79}
Cohen, M., Kuhi, L.V., 1979, ApJS 41, 743
\bibitem[1994]{dam94}
Damiani, F., Micela, G., Sciortino, S., et al., 1994, ApJ 436, 807
\bibitem[1999]{gal99}
Galli, D., Lizano, S., Li, Z.Y., et al., 1999, ApJ 521, 630
\bibitem[1993]{G93}
Geiss, J., 1993, In Origin and Evolution of the Elements, Eds. Prantzos, N.,
Vangioni-Flam, E., Cass, M., Cambridge Univ. Press, p. 89 
\bibitem[1997]{hanson}
Hanson, M.M., Howarth, I.D., Conti, P.S., 1997, ApJ 489, 698
\bibitem[2000]{Henn00}
Henning, Th., Schreyer, K., Launhardt, R., et al., 2000, A\&A 353, 211
\bibitem[1992]{hill92}
Hillenbrand, L.A., 1992, ApJ 397, 613
\bibitem[1997]{HoCh97}
Hofner, P., Churchwell, E. 1997, ApJL 486, L39
\bibitem[1981]{larson}
Larson, R.B., 1981, MNRAS 145, 271
\bibitem[1996]{McL96}
McLaughlin, D.E., Pudritz, R.E., 1996, ApJ 469, 194
\bibitem[1997]{McL97}
McLaughlin, D.E., Pudritz, R.E., 1997, ApJ 476, 750
\bibitem[1999]{oso99}
Osorio, M., Lizano, S., D'Alessio, P., 1999, ApJ 525, 808
\bibitem[1993]{PS93}
Palla, F., Stahler, S.W., 1993, ApJ 418, 414
\bibitem[1999]{PS99}
Palla, F., Stahler, S.W., 1999, ApJ 525, 772
\bibitem[1994]{Poll94}
Pollack J.B., Hollenbach D., Beckwith S., et al., 1994, ApJ 421, 615
\bibitem[1992]{schaller}
Schaller, G., Schaerer, D., Meynet, G., et al., 1992, A\&AS 96, 269
\bibitem[1996]{she96}
Shepherd, D.S., Churchwell, E., 1996, ApJ 472, 225
\bibitem[1988]{Sta88}
Stahler, S.W. 1988, ApJ 332, 804
\bibitem[2000]{Sta98}
Stahler, S.W., Palla, F., Ho, P.T., 2000, in Protostars and
Planets IV, Eds. V. Mannings et al., Univ. of Arizona Press,
Tucson.
\bibitem[1981]{Sta81}
Stahler, S.W., Shu, F.H., Taam, R.E., 1981, ApJ 248, 727
\bibitem[1980a]{Sta80a}
Stahler, S.W., Shu, F.H., Taam, R.E., 1980a, ApJ 241, 637
\bibitem[1980b]{Sta80b}
Stahler, S.W., Shu, F.H., Taam, R.E., 1980b, ApJ 242, 226
\bibitem[1990]{the90}
Th\'e, P.S., de Winter, D., Feinstein, A., et al. 1990, A\&AS 82, 319
\bibitem[1999]{Tout99}
Tout, C.A., Livio M., Bonnell, I.A. 1999, MNRAS 310, 360
\bibitem[1997]{win97}
de Winter, D., Koulis, C., Th\'e, P.S., et al., 1997, A\&AS 121, 223 
\bibitem[1987]{WC87}
Wolfire, M.G., Cassinelli, J.P., 1987, ApJ 319, 850
\end{thebibliography}
\end{document}